\documentclass[prd,a4paper,preprintnumbers,amsmath,amssymb,showkeys,showpacs]{revtex4}
\usepackage{epsfig}
\begin{document}

\markboth{}
{\textit{D. A. Fagundes, M. J. Menon, P. V. R. G. Silva}}

\title{Preliminary Results on the Empirical Applicability of the
Tsallis Distribution in Elastic Hadron Scattering}

\author{D. A. Fagundes, M. J. Menon, P. V. R. G. Silva}

\affiliation{Universidade Estadual de Campinas - UNICAMP\\
Instituto de F\'{\i}sica Gleb Wataghin \\
13083-859 Campinas, SP, Brazil \\
fagundes@ifi.unicamp.br, menon@ifi.unicamp.br, precchia@ifi.unicamp.br}

\begin{abstract}
We show that the proton-proton elastic differential cross section data
at dip position and beyond
can be quite well described by a parametrization based on the Tsallis
distribution, with only five free fit parameters. 
Extrapolation of the results obtained at 7 TeV to large
momentum transfer, suggests that hadrons may not behave as
a black-disk at the asymptotic energy region.
\end{abstract}


\maketitle

\centerline{\textit{Contribution to XII Hadron Physics, Bento Gon\c 
calves - RS,
Brazil, 22-27th April 2012}}

\centerline{\textit{To be published in AIP Proc. Conf.}}

\vspace*{15pt}

The theoretical description of the \textit{elastic} hadron-hadron 
differential cross section still constitutes a hard challenge for QCD. 
Experimental information available on
proton-proton ($pp$) scattering at 19 - 60 GeV comprises \textit{eleven to twelve decades of
experimental data}, which turns out very difficult any phenomenological
\cite{review} or even empirical \cite{empirical} description of this quantity.
Moreover, recent results obtained by the TOTEM Collaboration
at 7 TeV have indicated that
representative-model extrapolations to this energy are not consistent with the bulk of the experimental
information, in special at large values of the momentum transfer \cite{totem1,totem2}.

On the other hand,
the Tsallis distribution \cite{tsallis} has been widely used in the description
of the transverse momentum spectra of identified particles 
\cite{wilk,cleymans,de}.
A striking
feature is the fact that these spectra have
similar shapes as those characteristic of the elastic differential cross
sections at and beyond the dip position. Based on this similarity,
we have developed an empirical analysis of the $pp$ differential
cross section data with a parametrization based on the Tsallis
distribution for the intermediate and high momentum transfer squared regions,
together with  two exponential
functions for the low momentum region (the diffraction peak). 

Different forms of the Tsallis distribution \cite{tsallis} have been used in the literature 
and several fundamental aspects
are discussed, for example, in \cite{wilk,cleymans,de} 
and references therein.
For our purpose we shall consider the version that
has been used by the experimental collaborations (STAR, PHENIX, ALICE, ATLAS and CMS),
which is expressed by
\begin{eqnarray}
\frac{d^2 N}{dp_{T} dy} = \frac{dN}{dy}
\frac{(n-1)(n-2)}{nT[nT + m_0 (n-2)]}\,
p_T
\left[ 1 + \frac{\sqrt{p_T^2 + m_0^2} - m_0}{nT}\right]^{-n},
\end{eqnarray}
where $p_T$ is the transverse momentum, $dN/dy$ the yield per unity
rapidity $y$, $m_0$ the fixed hadron mass and $n$ and $T$ are shape parameters.

The main ingredient in this
communication is to show that an analytical parametrization based on the above distribution
can give quite good descriptions of the $pp$ differential cross section data at and beyond the dip
position. However, in order to get a global description 
of the available data,
an \textit{instrumental sum of two exponential} are also considered for the region below the dip position
(the diffraction peak region). For a matter of notation in a strictly empirical context, we shall 
denote these two forms by the regions they are intended for, namely: \textit{Diffraction Peak} ($DP$) contribution 
and \textit{beyond the Diffraction
Peak} ($bDP$) contribution.
Specifically, we express the differential cross section in terms of the 
\textit{momentum transfer squared}, $q^2$, as a sum of two terms,
\begin{eqnarray}
\frac{d\sigma}{dq^2} = \left.{\frac{d\sigma}{dq^2}}\right|_{DP} +  \left.{\frac{d\sigma}{dq^2}}\right|_{bDP}.
\end{eqnarray}
The first one is parametrized as
\begin{eqnarray}
\left.{\frac{d\sigma}{dq^2}}\right|_{DP} = a_1 e^{-b_1 q^2} + a_2 e^{-b_2 q^2},
\end{eqnarray}
where $a_i$, $b_i$, $i = 1, 2$ are free fit parameters. The second one is based on the
Tsallis distribution shifted from the origin, 
\begin{eqnarray}
\left.{\frac{d\sigma}{dq^2}}\right|_{bDP} = 
\alpha \frac{(\gamma-1)(\gamma-2)}{\gamma \beta[\gamma \beta + \delta (\gamma-2)]}
[q^2 - q_0^2]
\left[ 1 + \frac{\sqrt{ [q^2 - q_0^2]^2 + \delta^2} - \delta}{\gamma \beta} \right]^{-\gamma},
\end{eqnarray}
where $\alpha$, $\beta$, $\gamma$, $\delta$ and $q_0^2$
are free fit parameters.

As in previous empirical investigation \cite{empirical},  we shall base our analysis 
on six
sets of $pp$ elastic scattering data, at $\sqrt{s}$ = 19.4 GeV,
23.5, 30.7, 44.7, 52.8 and 62.5 GeV.
In addition, we make use of the empirical result that at the ISR energy region (23.5 - 62.5 GeV)
the differential cross section data above $q^2 \sim$ 4 GeV$^2$ do not depend on 
the energy.
This fact allows the inclusion in each set (ISR) the data at 27.4 GeV (from Fermilab), covering the
large momentum transfer region: 5.5 $\leq q^2 \leq$ 
14.2 GeV$^2$ (see \'Avila and Menon in \cite{empirical} for a quantitative discussion on this respect).
Only the statistical uncertainties of the experimental data are taken into account.
Here, we shall also consider the new results on elastic $pp$ scattering at 7 TeV, recently obtained 
by the TOTEM Collaboration at the LHC.
However, since
the data points (numerical values) are not yet available in published format,
we have used the applicative Plot Digitizer to extract the points and uncertainties from the
plots in \cite{totem1} and \cite{totem2}.
The data reductions have been performed with the objects of the class TMinuit of ROOT Framework. 
First, the two regions have been
treated separately in order to infer start values of the parameters in a global
final fit and in that case we have included in the fit code a step-function
of $q^2 - q_0^2$,
before the bDP parametrization (4).
Typical fit results (curves) are presented in Figure 1.
The values of the fit parameters at 7 TeV are:
$a_1$ = 505.4 $\pm$1.4 mbGeV$^{-2}$,
$b_1$ = 20.429 $\pm$0.029 GeV$^{-2}$,
$\alpha$ = 1.1321 $\pm$0.0094 $\times 10^{-2}$ mb,
$\beta$ = 0.1339 $\pm$ 0.0056 GeV$^2$,
$\gamma$ =  19.6 $\pm$ 3.1,
$\delta$ = 0.216 $\pm$ 0.029 GeV$^2$ and
$q_0^2$ = 0.5015 $\pm$ 0.0029 GeV$^2$
($\chi^2$/DOF = 1.65 for 115 DOF).

\begin{figure}
\includegraphics*[width=5.9cm,height=7.5cm]{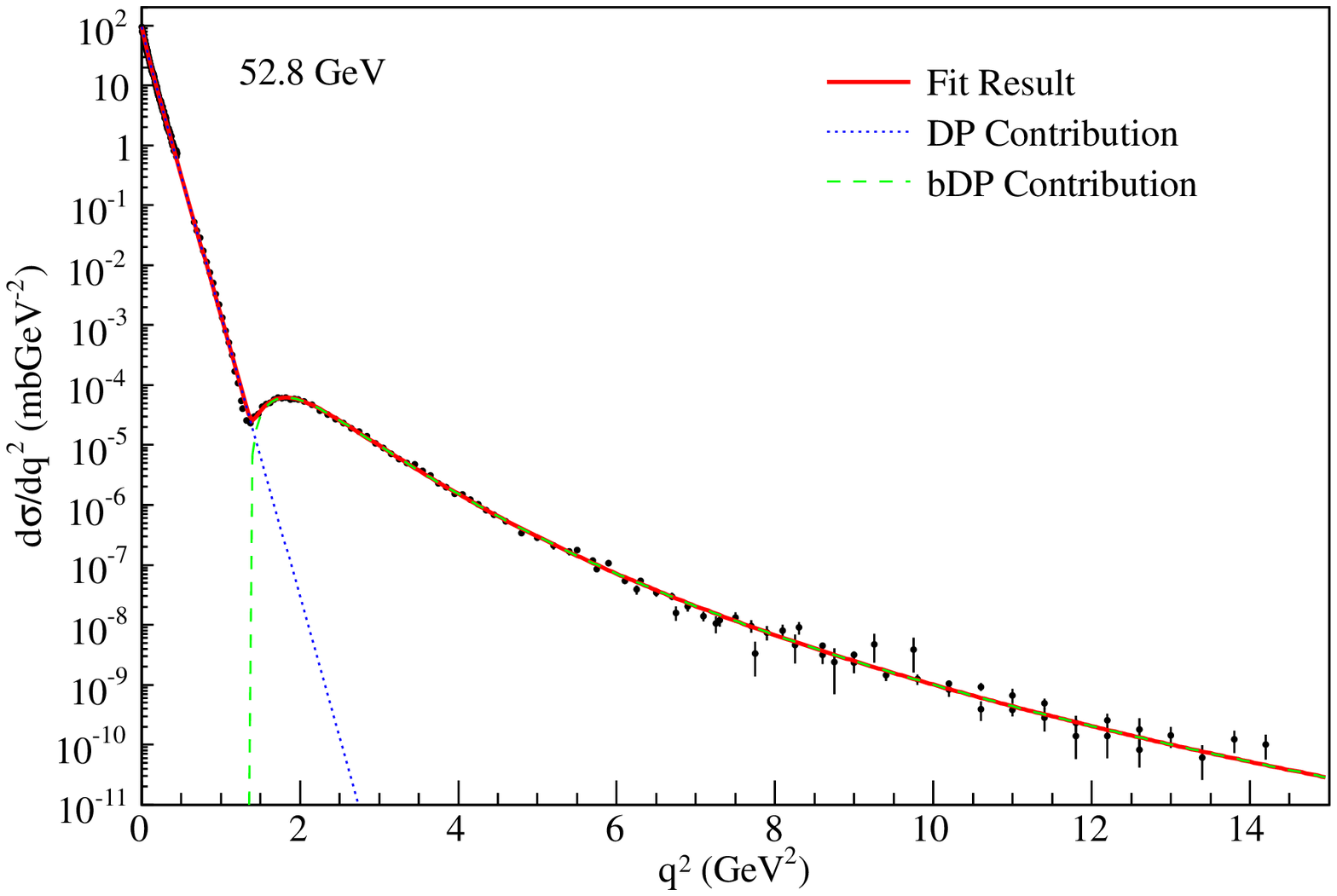}\hspace*{-0.5cm}
\includegraphics*[width=5.9cm,height=7.5cm]{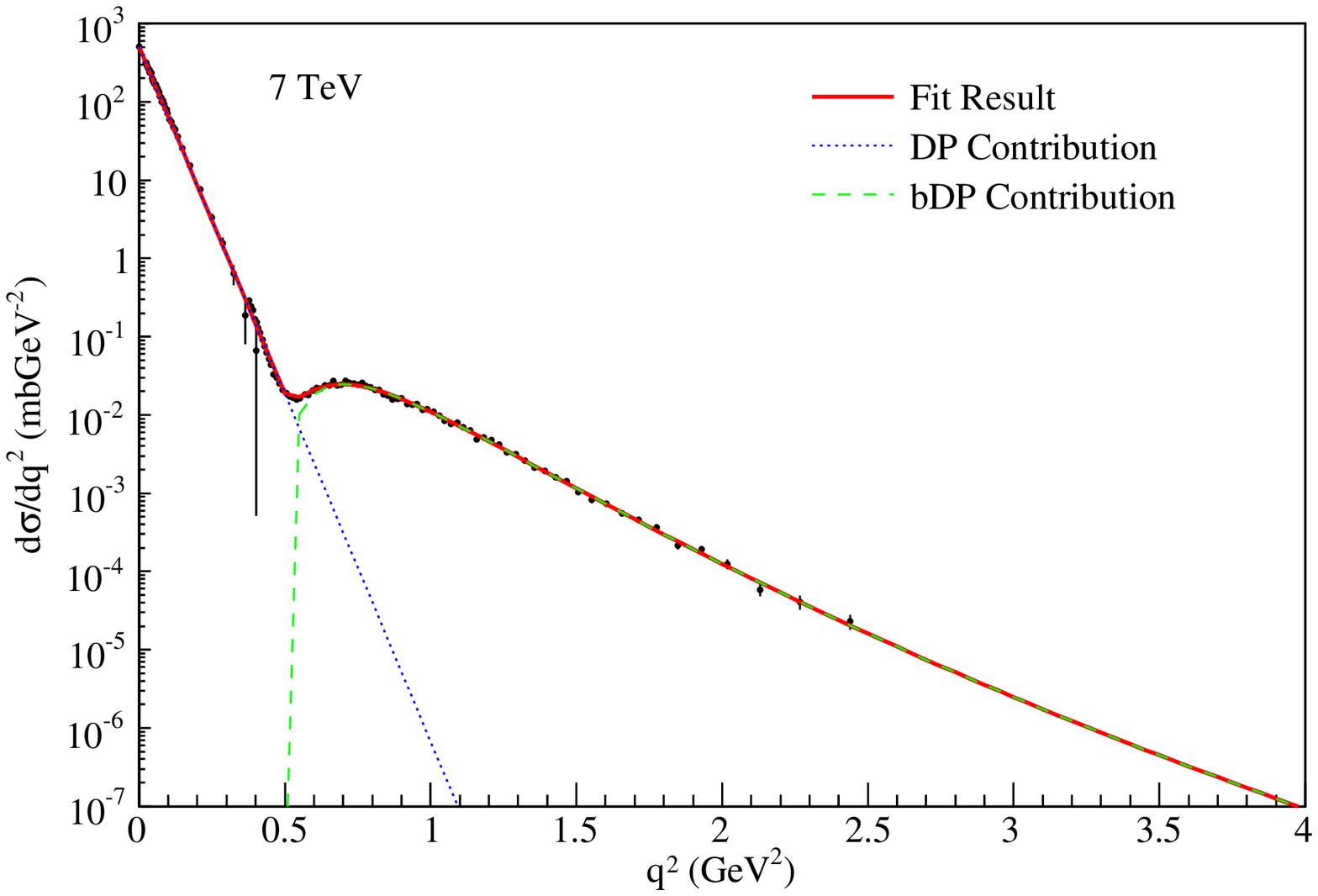}\hspace*{-0.5cm}
\includegraphics*[width=5.9cm,height=7.5cm]{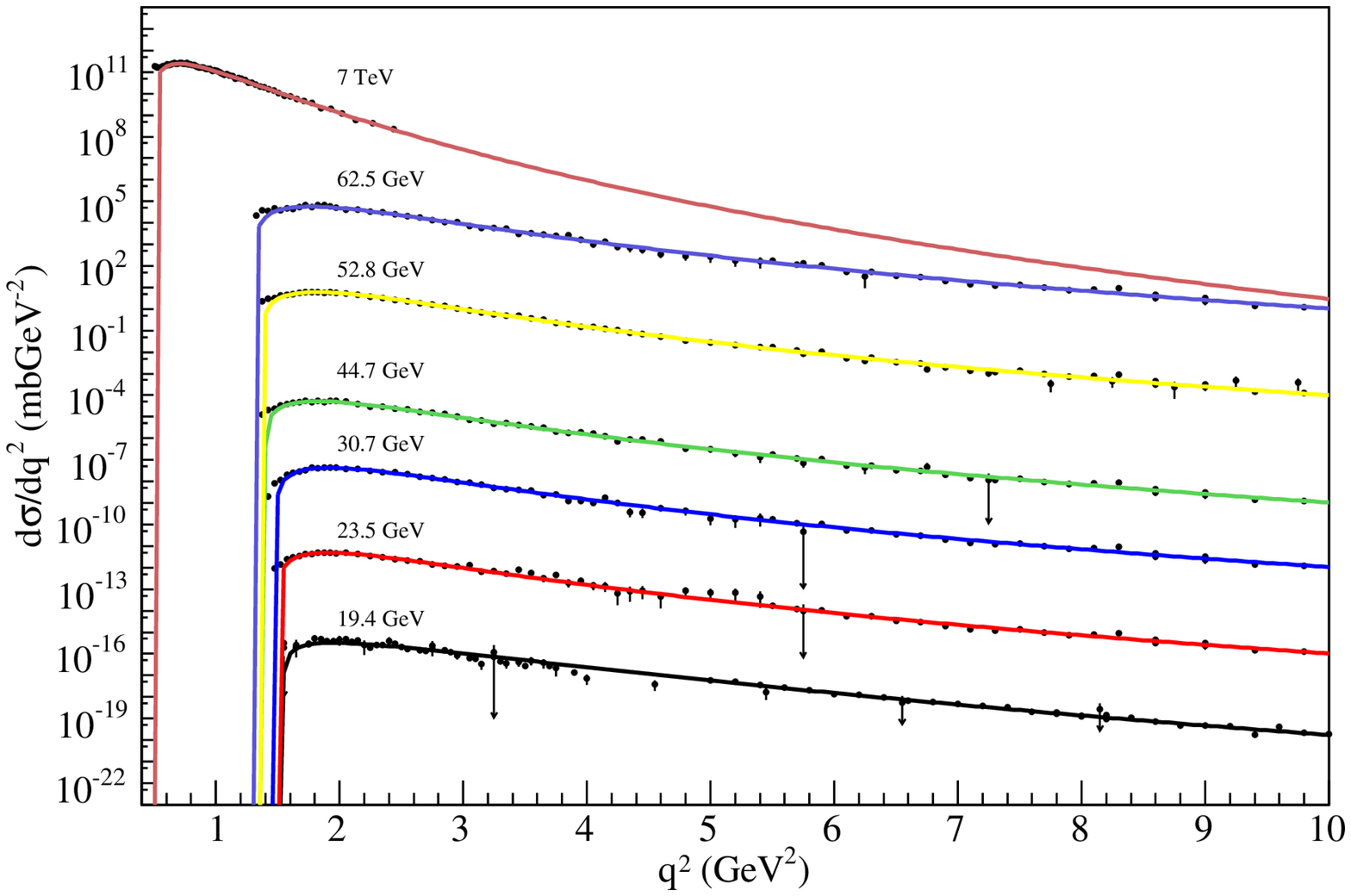}
\caption{Left and center: Typical fit results and contributions to the differential cross section from the
\textit{Diffraction Peak} (DP) parametrization (3) and \textit{beyond the Diffraction Peak} (bDP)
parametrization (4), together with the experimental data (52.8 GeV and 27.4 GeV) and extracted points
(7 TeV). Right: All results at the dip position and beyond with the bDP contribution, Eq. (4).
Curves and data have been multiplied by factors of 10$^{\pm 2}$}
\end{figure}

We conclude that parametrization (2-4) leads to good descriptions of all the $pp$
differential cross section data available at $\sqrt{s} \geq$ 19.4 GeV.
A novel result  concerns
the monotonic  decrease of the differential cross section (without any oscillation) at the \textit{deep
elastic scattering} region, $q^2$ above $\sim$ 4 GeV$^2$ \cite{islam},  even at 7 TeV
(where the slope, with a small positive curvature, is nearly two times the value observed at the ISR region,
as illustrated in Figure 1, right).
Therefore, this pattern suggests a scenario in disagreement with the black-disk as an
asymptotic limit. 
Another novel aspect, concerns the suggestion of a possible
nonextensive thermal statistics
interpretation of the dip-bump and deep elastic scattering regions.
Further tests and investigation are in course.

%% BACKMATTER
%%%%%%%%%%%%%%%%%%%%%%%%%%%%%%%%%%%%%%%%%%%%%%%%

\begin{acknowledgments}
We are thankful to 
J. Takahashi and D.D. Chinellato for discussions.
Research supported by FAPESP 
(Contracts Nos. 11/15016-4, 11/00505-0, 09/50180-0).

\end{acknowledgments}

\end{document}